\documentclass{optica-article}

\journal{opticajournal} % for journals or Optica Open

\articletype{Research Article}

\usepackage[version=4]{mhchem}
\usepackage{booktabs}
\usepackage{makecell}
\usepackage{lineno}
\usepackage{dcolumn}

\newcommand{\op}[1]{\hat{#1}}

\begin{document}

\title{Squeezing via self-induced transparency in mercury-filled photonic crystal fibers}

\author{M.~S.~Najafabadi,\authormark{1,*} J.~F.~Corney,\authormark{2} L. L. S\'anchez-Soto,\authormark{1,3} N. Y.~Joly,\authormark{4,1} and G. Leuchs\authormark{1,4}}

\address{\authormark{1}Max-Planck-Institut f\"{u}r die Physik des Lichts, 91058~Erlangen, Germany\\
\authormark{2}School of Mathematics and Physics, University of Queensland, Brisbane, Queensland 4072, Australia\\
\authormark{3}Departamento de \'Optica, Facultad de F\'{\i}sica, Universidad Complutense, 28040~Madrid, Spain\\
\authormark{4}Institut f\"{u}r Optik, Information und Photonik,  Friedrich-Alexander-Universit\"{a}t Erlangen-N\"{u}rnberg, 91058~Erlangen, Germany}

\email{\authormark{*}mojdeh.shikhali-najafabadi@mpl.mpg.de} %% email address is required; see note below about the corresponding author designation

% use {asbstract*} to suppress the copyright line. Copyright information will be added in production

\begin{abstract*} 
We investigate the squeezing of ultrashort pulses using self-induced transparency in a mercury-filled hollow-core photonic crystal fiber. Our focus is on quadrature squeezing at low mercury vapor pressures, with atoms near resonance on the $\ce{^{3}{}D_{3}} \rightarrow 6\ce{^{3}{}P_{2}}$ transition. We vary the atomic density, thus the gas pressure (from  2.72 to {15.7~$\mu$bar)}, by adjusting the temperature (from 273~K to 303~K). Our results show that achieving squeezing at room temperature, considering both fermionic and bosonic mercury isotopes, requires ultrashort femtosecond pulses. We also determine the optimal detection length for squeezing at different pressures and temperatures.
\end{abstract*}

\section{Introduction}

Self-induced transparency (SIT) is a well-established phenomenon in nonlinear optics, where a coherent light pulse can propagate through a resonant medium without being absorbed, even though the medium would typically be opaque at that frequency. This effect was first reported by McCall and Hahn~\cite{McCall:1967aa,McCall:1969aa}, who, using a semiclassical approach, showed that a two-level medium becomes transparent to a pulse with an area {of $2\pi$~\cite{Allen:2012aa}.} Since then, SIT has been observed across various media (for comprehensive reviews, see~\cite{Lamb:1971aa,Slusher:1974aa,Poluektov:1975aa,Maimistov:1990aa}).    

While there are several popular methods for generating squeezed states, these processes often have limitations, such as high-power requirements and phase-matching sensitivity, which complicate their application. SIT offers a unique alternative by leveraging coherent atomic population oscillations, which do not require strong pump fields, reducing the need for high-intensity lasers. Additionally, SIT avoids strict phase-matching conditions, making it easier to implement. Thus, SIT solitons have emerged as promising candidates for generating squeezed states, overcoming many challenges posed by traditional methods~\cite{Watanabe:1989aa}.

However, experimental demonstrations in gases have remained elusive. To overcome these challenges, gas-filled hollow-core photonic crystal fibers (PCFs) have emerged as an alternative for squeezing generation. These fibers exploit the strong nonlinearity of atomic transitions alongside a tight transverse confinement of light~\cite{Russell:2006aa,Travers:2011aa}. A notable  advantage of this approach is the ability to fine-tune the nonlinearity by adjusting the gas pressure.

Despite their potential, gas-filled PCFs also pose certain challenges, particularly in filling them with a suitable gas~\cite{Ghenuche:2012aa}. For instance, alkali vapors, such as rubidium, tend to bond with and diffuse into the glass walls, complicating the process of maintaining a consistent pressure~\cite{Yu:2020aa}. Recent advancements, however, have demonstrated that it is possible to neatly fill hollow-core PCFs with rubidium in an adapted setup~\cite{Haupl:2022aa}.

Alternatively, mercury has been suggested as a filling gas~\cite{Vogl:2014aa}, owing to its high vapor pressure at room temperature, which prevents condensation on the glass. Moreover, mercury demonstrates significant optical nonlinearities in the ultraviolet and blue regions of the spectrum.

{Recent advancements in radiation-matter interaction, both theoretical and experimental, have been remarkable. The ultrastrong coupling regime has been realized in cavity and circuit electrodynamics, as well as in cavity optomechanics~\cite{Qin:2024aa}. However, its application to SIT remains unexplored, largely due to the reliance on semiclassical models, using linearization~\cite{Lai:1990aa,Lee:2009aa}  or back-propagation methods~\cite{Lai:1995aa} for studying quantum noise. A full quantum approach is therefore crucial for a deeper understanding of the underlying physics.} 

{To achieve this,} we adapt a method for studying the propagation of radiation within an optically pumped two-level system, incorporating both collisional and radiative damping effects~\cite{Drummond:1991aa}. The approach involves deriving a set of stochastic $c$-number differential equations that are equivalent to the Heisenberg operator equations. This is done using the positive-$P$ representation~\cite{Drummond:1980aa}, which provides a probabilistic framework wherein stochastic averages correspond to normally ordered correlations. A significant advantage of this technique is its ability to yield numerically solvable equations while retaining key aspects that represent the quantum nature of the field.  

{Leveraging this powerful tool, we undertake an in-depth exploration of the feasibility of using mercury-filled PCFs to generate pulsed squeezed states. Aimed at supporting experimentalists working with mercury atoms in SIT experiments, this work incorporates the full atomic structure and transition dynamics, offering a comprehensive and realistic framework to guide and advance future research.} 

The structure of this paper is as follows: {In Sec.~\ref{sec:model}, we introduce the model Hamiltonian and examine how quantum noise arises from both damping and nonlinearities. Building on our earlier work~\cite{najafabadi:2024aa}, which focused on two-level transitions, we now extend this approach by incorporating all hyperfine and dipole-allowed transitions for both bosonic and fermionic isotopes of mercury. These additional transitions are essential for accurately modeling the behavior of mercury atoms in SIT experiments. In Sec.~\ref{sec:simul}, we explore the resulting dynamics by numerically solving the fully nonlinear stochastic differential equations derived from the positive-$P$ representation.} In Sec.~\ref{sec:res}, we present the main results of our model. First, we show that for a sample containing only the isotope $\ce{^{202}Hg}$, the squeezing is several dB higher than in a sample containing all isotopes at 273 K. We also analyze the effect of detuning on squeezing for $\ce{^{202}Hg}$ at 273 and 293 K, demonstrating that detuning enhances squeezing at these temperatures. Finally, we identify the optimal squeezing across a range of temperatures and determine the corresponding detection lengths for each case. Our conclusions are summarized in Sec.~\ref{sec:Conc}.

\section{Model Hamiltonian}
\label{sec:model}

To make this paper as self-contained as possible, we will briefly review the model we use. Based on the ideas in Ref.~\cite{Drummond:1991aa}, we present a Hamiltonian that describes the interaction between an ensemble of two-level atoms and a single mode of the radiation field. In the rotating-wave and dipole approximations, this Hamiltonian reads
\begin{equation}
\label{eq:H}
    \op{H} = \op{H}_{\mathrm{A}}+\op{H}_{\mathrm{F}}+ \op{H}_{\mathrm{B}} + \op{H}_{\mathrm{FB}} + \op{H}_{\mathrm{AB}} + \op{H}_{\mathrm{AF}} 
\end{equation}
where
\begin{equation}
\begin{aligned}
    & \hat{H}_\mathrm{A}  = \frac{1}{2} \sum_{\mu} \hbar \omega_{\mu} \hat{\sigma}^{3}_{\mu},\\
    & \hat{H}_\mathrm{F}  = \sum_{k} \hbar \omega_{k} \, \hat{a}^{\dag}_{k} \hat{a}_{k},\\
    & \hat{H}_\mathrm{B}  =\hat{H}^{a}+\hat{H}^{\sigma}+ \hat{H}^{z},\\
    & \hat{H}_{\mathrm{AF}}  =\hbar \sum_{k} \sum_{\mu} (g\hat{a}^{\dag}_{k}\hat{\sigma}^{-}_{\mu} e^{-ik z_{\mu}} + \mathrm{H.c.}), 
\end{aligned}
\end{equation}
In this model, $\hat{H}_{\mathrm{A}}$ represents the free Hamiltonian of the atoms, where $\omega_{\mu}$ is the resonant frequency of the $\mu$th atom, described using standard Pauli operators~\cite{Gardiner:2004aa}:
\begin{equation}
 {   \hat{\sigma}_\mu^+ = |e\rangle_{\mu} \langle g |_{\mu} ,
    \quad
    \hat{\sigma}_\mu^- = |g\rangle_{\mu} \langle e |_{\mu} ,
    \quad
    \hat{\sigma}^{3}_{\mu} = |e\rangle_{\mu} \langle e |_{\mu} - |g \rangle_{\mu} \langle g |_{\mu} ,}
\end{equation}
{with  $|e\rangle_{\mu} $ and $|g \rangle_{\mu}$ being the excited and ground state, respectively}.  Similarly, $\hat{H}_{\mathrm{F}}$ represents the free Hamiltonian of the field modes traveling through the fiber, with each mode characterized by a frequency $\omega_{k}$ and annihilation operator $\hat{a}_{k}$ (for a single polarization). 

{The term $\hat{H}_{\mathrm{B}}$ is the free Hamiltonian of the reservoirs, which includes contributions from the field modes ($\hat{H}^{a}$), atomic dipoles ($\hat{H}^{\sigma}$), and collisions ($\hat{H}^{z}$).  The interaction between the field and atomic dipoles is described by $\hat{H}_{\mathrm{AF}}$ , with a dipole-field coupling constant $g$, assumed to be the same for all atoms and independent of frequency and wave vector.  The Hamiltonian $\hat{H}_{\mathrm{AB}}$ accounts for the interaction between the atomic reservoirs and atoms, while $\hat{H}_{\mathrm{FB}}$ describes the interaction of the background reservoir with the radiation field.} 

{{The system's evolution can be studied through the master equation. By tracing out the reservoir variables and applying the standard Markov approximation~\cite{Gardiner:2004aa}, we obtain:
\begin{align}
\label{Eq:master_Eq}
  \frac{d \hat{\varrho}}{dt} = \frac{1}{i\hbar} [ \hat{H} , \hat{\varrho} ] + \hat{\mathcal{L}}_{\mathrm{AB}} [ \hat{\varrho} ] + \hat{\mathcal{L}}_{\mathrm{FB}}[ \hat{\varrho}] \, , 
\end{align}
where $\hat{\varrho}$ is the density matrix of the system. The Lindblad superoperators $\hat{\mathcal{L}}_{\mathrm{AB}}$ and $\hat{\mathcal{L}}_{\mathrm{FB}}$ describe relaxation processes in the atomic and field variables, respectively, taking the following form:
%\begin{widetext}
\begin{equation}
\begin{aligned}
   \hat{\mathcal{L}}_{\mathrm{AB}} [\hat{\varrho}] & = 
    \sum_{\mu}  \tfrac{1}{2} W_{21}   ( [ \hat{\sigma}^{-}_{\mu} \hat{\varrho}, \hat{\sigma}^{+}_{\mu} ] + [ \hat{\sigma}_{\mu}^{-}, \hat{\varrho}\sigma_{\mu}^{+} ] ) 
    + \tfrac{1}{2}W_{12} ( [ \sigma_{\mu}^{+}\hat{\varrho}, \hat{\sigma}_{\mu}^{-} ]  + [ \sigma_{\mu}^{+}, \hat{\varrho}\hat{\sigma}_{\mu}^{-} ] )  + \tfrac{1}{4} \gamma_{p}  ( [ \hat{\sigma}_{\mu}^{3}, \hat{\varrho}\hat{\sigma}_{\mu}^{3} ] + [ \hat{\sigma}_{\mu}^{3}\hat{\varrho}, \hat{\sigma}_{\mu}^{3} ] ) \, , \\
      \hat{\mathcal{L}}_{\mathrm{FB}} [\hat{\varrho}] & = \tfrac{1}{2} c \kappa  \sum_{k}
     ( 1+ \bar{n}) ( [\hat{a}_{k}  \hat{\varrho}, \hat{a}_{k}^{\dag} ] + [ \hat{a}_{k}, \hat{\varrho}\hat{a}_{k}^{\dag} ]) + \bar{n}([ \hat{a}_{k}^{\dag}\hat{\varrho}, \hat{a}_{k}] + [ \hat{a}_{k}^{\dag}, \hat{\varrho}\hat{a}_{k} ] ). 
  \end{aligned}
\end{equation}
%\end{widetext}
Here, $W_{21}$ is the relaxation rate from the excited to the ground state, $W_{12}$ is the incoherent pumping rate, and $\gamma_{p}=3\gamma_{0}$ is the pure dephasing rate. For the field, $\kappa$ is the absorption rate during the propagation within the medium and 
\begin{equation}
\label{photonnumb}
\bar{n}= \frac{1}{\exp\left (\frac{\hbar \omega_{0}}{k_{B}T_{f}}\right )-1}
\end{equation} 
is the mean equilibrium photon number in each reservoir mode of interest with $T_{f}$ to be temperature of the field background reservoir. 

If we consider the thermal temperature of the radiative reservoir for the atoms to be $T_{a}$, then:
\begin{align}
    W_{21}=\gamma_{0} (1+ \bar{n}_{a} )\, ,  \qquad  \qquad   W_{12} = \gamma_{0} \bar{n}_{a} \, ,
\end{align}
with photon occupation number $\bar{n}_{a}$ given by \eqref{photonnumb} with temperature $T_{a}$.  

The damping rates are
\begin{equation}
   \gamma_{\|} = W_{12} +W_{21} \, , \qquad \qquad 
    \gamma_{\perp} = \gamma_{p}+\tfrac{1}{2} \gamma_{\|}  \, .  
\end{equation}
These coefficients $\gamma_{\|}$ and $\gamma_{\perp}$ correspond to two different damping mechanisms; namely, longitudinal (population decay) and transverse (dephasing). }

We assume that only one transverse mode of the optical waveguide is relevant, meaning that other transverse modes are either unsupported by the waveguide or not excited during propagation. Additionally, the transverse mode profile $u({\bf r}_\perp)$  remains uniform along the length of the fiber. The relevant electromagnetic modes are those with wave vectors aligned along the waveguide axis (assumed to be the $z$ axis). Therefore, we can define the optical field (scaled to the Rabi frequency) at position  $z_l$ as
\begin{align}
\hat{\Omega} (z_l) \equiv 2ig \sum_m  e^{i k_m z_{l}} \hat{a}_m \, .
\end{align}
Here,  $\hat{a}_m$  are the annihilation operators for modes with frequency  $\omega_m$ and corresponding wavenumbers $k_m = m \Delta k$, where $\Delta k = 2\pi/L$ and $L$ is the quantization length (i.e., length of fibre).  

We define the atomic field operators by adding together the operators for individual atoms in a given spatial cell at ${\bf r}_j$ and within the frequency band centred at $\omega_m$:
\begin{equation}
\begin{aligned}
 \hat{R}^{3} (\mathbf{r}_j, \omega_m) & \equiv \frac{1}{N_{jm}} \sum_{n}^{N_{jm}} \hat{\sigma}_{jmn}^{3} \, , \\ 
 \hat{R}^{\pm}(\mathbf{r}_j, \omega_m) & \equiv \frac{2}{N_{jm}} \sum_{n}^{N_{jm}} \hat{\sigma}_{jmn}^{\pm}; 
\end{aligned}
\end{equation}
where the atomic index $\mu =(j,m,n)$ has been expanded so that we can sum over only those atoms satisfying the above conditions.  Note that the number of atoms in each spatio-frequency cell can be written as $N_{jm} = \rho(\mathbf{r}_j, \omega_m)  \Delta V \Delta \omega$, where  $\rho(\mathbf{r}_j, \omega_m)$ is the density of resonant atoms in $\mathbf{r}_{j}$ and a certain frequency $\omega_{m}$.

Using standard methods one can derive the corresponding master equation. However, its direct numerical integration is extremely difficult. For that reason, our strategy is to derive suitable equations of motion in phase space. We use the positive $P$ approach~\cite{Drummond:1980aa}, which is a normally ordered operator representation that identifies the moments of $\hat{\varrho}$ with the corresponding $c$-number moments of a positive $P$ distribution. 

In this approach, we have a mapping $\hat{\Omega} \leftrightarrow \Omega$, $\hat{\Omega}^{\dag} \leftrightarrow \Omega^{\dag}$, $\hat{R}^{\pm} \leftrightarrow R^{\pm}$, $\hat{R}^{3} \leftrightarrow R^{3}$ and, following the standard procedures, the master equation can then be transformed into an equivalent Fokker-Planck equation for $P(\Omega, \Omega^{\dag}, R^{-}, R^{+}, R^{3})$. This equation is valid only when the distribution $P(\Omega, \Omega^{\dag}, R^{-}, R^{+}, R^{3})$ vanishes sufficiently rapidly at the boundaries. In practical applications, it is usually the case that the damping terms provide a strong bound at infinity on the distribution function~\cite{Gilchrist:1997aa}.

In terms of these variables, and in the limit of large number of atoms, { we derive stochastic equations similar to those in~\cite{najafabadi:2024aa}, adapted to account for the presence of multiple isotopes in our mercury gas sample. These adjustments yield the following set of equations, which serve as the foundation for our simulations:
\begin{equation}
 \begin{aligned}
   \label{field_R_6}
  & \left ( \frac{\partial}{\partial  z} +\frac{1}{c}\frac{\partial}{\partial t} \right ) \Omega(t, z)  = -\frac{1}{2}\kappa \Omega(t, z) + \sum_{\alpha=1} \sum_{\beta=1} G_{\alpha}  \int \rho_{\alpha}( \mathbf{r},\omega)R^{-}_{\alpha \beta}(t, \mathbf{r},\omega) \, d\mathbf{r}_\perp d\omega 
  +F^{\Omega}(t, z), & \\
&\frac{\partial}{\partial t}R^{-}_{\alpha \beta}(t, \mathbf{r},\omega)  = - [\gamma_{\perp} + i(\omega -\omega_{0}) ] R^{-}_{\alpha \beta}(t, \mathbf{r},\omega)  + u (\mathbf{r}_\perp) \Omega(t, z)R^{3}_{\alpha \beta}(t, \mathbf{r},\omega)+F^{R}(t,  \mathbf{r}, \omega),& \\
 &\frac{\partial}{\partial t} R^{3}_{\alpha \beta }(t, \mathbf{r},\omega) = - \gamma_{\|} [R^{3}_{ij} (t, \mathbf{r},\omega)-\sigma^{SS} ] -\frac{1}{2} [ u (\mathbf{r}_\perp) \Omega(t, z)R^{+}_{\alpha \beta}(t, \mathrm{r},\omega) + u^{\ast} (\mathbf{r}_\perp) \Omega^{\dag}(t, z) R^{-}_{\alpha \beta}(t, \mathbf{r},\omega) ]+ F^{z}(t,  \mathbf{r}, \omega),&  
 \end{aligned}
 \end{equation}
where 
\begin{equation}
\sigma^{SS} = \frac{W_{12}-W_{21}}{W_{12}+W_{21}} \,,
%\qquad  
%G_i=\frac{L g_i^{2}}{c} \, ,
\end{equation}
and $G_{\alpha}$ and $\rho_{\alpha}$ are the atomic-field interaction coefficient and the atomic density of $\alpha$th isotope, respectively, and are given by $\rho_{\alpha}(z, \omega)=A_{\alpha}~\rho_{\text{tot}}(z)  f_{\alpha}(\omega)$ where $A_{\alpha}$ is the abundance of isotope, $\rho_{\text{tot}}$ is the total atomic density, $f_{\alpha}(\omega)$ determines the lineshape of the atoms at the given temperature. ${R_{\alpha \beta}}$ represents the polarization vector of $\alpha_\mathrm{th}$ isotope for $\beta_\mathrm{th}$ transition, which in the fermionic case, $\sum_{\beta}$ is over all dipole-allowed transitions.
} 
Equations~\eqref{field_R_6} are identical with the usual semiclassical equations for the slowly varying envelope fields~\cite{Lax:1966aa,Haken:1966aa}, except for the presence of the Langevin terms $F$ that describe quantum fluctuations and depend on the bath and nonlinear atom-field coupling, and are expressed as:
{
    \begin{equation}
\begin{aligned}
     F^{\Omega}(t, z) & = 2\xi^{\alpha}(t, z) \sqrt{\sum_{\alpha} G_{\alpha}\kappa \overline{n}}= [F^{{\Omega}^{\dag}}(t, z)]^{\ast}, \\
      F^{R}(t,  \mathbf{r},\omega) & = \frac{1}{\sqrt{\rho_i(\mathbf{r}, \omega)}} \{ \xi^{J}(t, \mathbf{r}, \omega) \sqrt{ u (\mathbf{r}_\perp) \Omega R^{-}_{\alpha \beta}} + 2\xi^{P}(t, \mathbf{r},\omega)\sqrt{\gamma_{P}(R^{3}_{\alpha \beta}+1)} +2\xi^{o}(t, \mathbf{r}, \omega)\sqrt{W_{12}}\}, \\
       F^{R^{\dag}_{\alpha \beta }}(t,  \mathbf{r}, \omega) & = \frac{1}{\sqrt{\rho_{\alpha}(\mathbf{r}, \omega)}} \{ \xi^{J^{\dag}}(t,  \mathbf{r}, \omega) \sqrt{u^{\ast} (\mathbf{r}_\perp) \Omega^{\dag} R^{+}_{\alpha \beta}} +2\xi^{P\ast}(t,  \mathbf{r}, \omega)\sqrt{\gamma_{P}(R^{3}_{\alpha \beta}+1)}    +2\xi^{o\ast}(t,  \mathbf{r}, \omega)\sqrt{W_{12}} \} , \\
        F^{z}(t,  \mathbf{r}, \omega) & =\frac{1}{\sqrt{\rho_i(\mathbf{r}, \omega)}} \{ \xi^{z}(t,  \mathbf{r}, \omega)[2\gamma_{\parallel}(1 - \sigma^{SS}R^{3}_{\alpha \beta }) + (R^{-}_{\alpha \beta} u^{\ast} (\mathbf{r}_\perp) \Omega^{\dag} + R^{+}_{\alpha \beta } u (\mathbf{r}_\perp) \Omega)-2W_{12} R^{+}_{\alpha \beta}R^{-}_{\alpha \beta }]^{1/2} \\
       & \quad \quad \quad \quad \quad \quad-[\xi^{o}(t,  \mathbf{r},\omega) R^{+}_{\alpha \beta} + \xi^{o\ast}(t,  \mathbf{r},\omega) R^{-}_{\alpha \beta}] \sqrt{W_{12}} \} .
\end{aligned}
\end{equation}
}
The terms optical thermal noise  $\xi^{\alpha}(t, z)$, incoherent pumping noise $\xi^{o}(t, \mathbf{r},\omega)$ and collisional dephasing noise $\xi^{P}(t, \mathbf{r}, \omega)$ are complex, while photon-atom interaction noise $\xi^{J}(t, \mathbf{r},\omega), \xi^{J^{\dag}}(t, \mathbf{r},\omega)$ and $\xi^{z}(t, \mathbf{r},\omega)$ are real. These noise terms are $\delta$ correlated.  { Since the equations are derived through a normally ordered representation, there are bath noise terms associated with dephasing ($\gamma_p$) and gain ($W_{12}$ due to incoherent pumping from the ground to the excited state), and losses ($W_{21}$ due to relaxation rate from the excited state to the ground state)}.  Furthermore, the gain noise is only present at finite temperatures.  In addition to the bath noise, the positive-$P$ method has noise associated with the atom-field coupling, which is present even for unitary evolution and corresponds in some sense to shot-noise effects in the atom-light interaction.

\section{Simulation}
\label{sec:simul}

We investigate theoretically the squeezing generated due to SIT solitons using mercury atoms, whose advantages have been already discussed in the Introduction. The atomic mercury contains seven isotopes with more than 5\% abundance, including five bosonic { ($\ce{^{196}Hg}$, $\ce{^{198}Hg}$, $\ce{^{200}Hg}$, $\ce{^{202}Hg}$ and $\ce{^{204}Hg}$)} and two fermionic isotopes {$\ce{^{199}Hg}$ and $\ce{^{201}Hg}$} that are stable. { Table~\ref{tab:sample} shows the abundancy of the isotopes.} The fermions have low nuclear spin, simplifying the complexity of the hyperfine structure~\cite{mejri2011ultraviolet, villwock2011magneto}. Since bosonic and fermionic isotopes exhibit different dynamics, we will initially focus on the bosonic isotope $\ce{^{202}Hg}$, which has the highest abundance. The effect of other isotopes will be included afterwards in the simulation. 
\begin{table}[b]
    \centering
    \begin{tabular}{ccc}
        \hline
        Isotope & Abundance ($\%$)\\
        \hline
        196   & 0.15     \\
        198   & 10.02     \\
        200 & 23.13 \\
        202 & 29.80 \\
        204 & 6.85 \\
        199 & 16.84 \\
        201 & 13.22 \\
        \hline
    \end{tabular}
    \caption{ The abundances of the different isotopes are given in the table.}
    \label{tab:sample}
\end{table}
In this case, we identify the $6\ce{^{3}D_3} \rightarrow 6\ce{^{3}P_2}$  as the most suitable transition  since it is associated with metastable states and can exhibit long coherence times (see Fig.~\ref{fig:Energy_diagram}). This is crucial for supporting long-lived soliton states. 

%%%%%%%%%%%%%%%%%%%%%%%%Mercury%%%%%%%%%%%%%%%%%%%%%%%%%%%%
\begin{figure}[t]
    \centering
    \includegraphics[width=0.85\columnwidth]{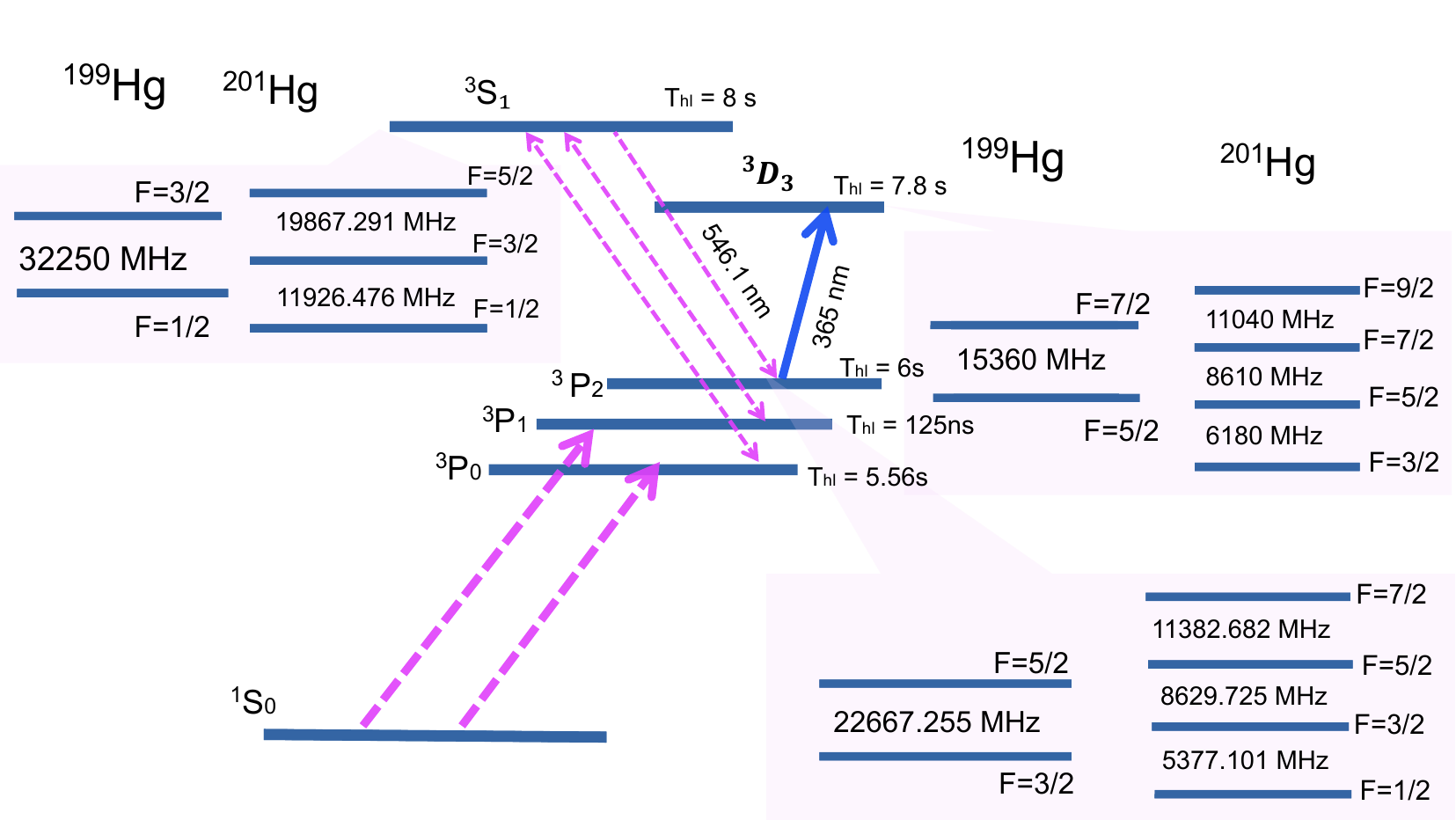}
    \caption{The relevant energy levels for neutral mercury (hyperfine structure for the fermions is plotted). The suitable transition for the SIT is considered to be $6\ce{^{3}_{}D_{3}}  \rightarrow {6\ce{^{3}_{}P_{2}}}$ transition. }
    \label{fig:Energy_diagram}
\end{figure}
%%%%%%%%%%%%%%%%%%%%%%%%%%%%%%%%%%%%%%%%%%%%%%%%%%%%%%%%%%%%%%%%%%%

We consider an optical hollow-core fiber that is ideally 50 mm long with a core diameter of 10~$\mu$m and filled with mercury vapor gas.  We assume the atomic profile to be {Lorentzian at zero Kelvin temperature}; however, at high temperatures, they have a Voigt profile. Experimentally, the fiber needs to be connected to a reservoir of Hg, from which one can change the density of atoms, thus the pressure through heating. This unavoidly yields a dead-volume which will be much larger than the volume of the fiber (20~$\mu$l for a 10~$\mu$m diameter, 50 mm long fiber). The dead volumes will impose the pressure in the channel connecting the reservoir and the fiber.

We assume the gas is homogeneously distributed throughout the PCF, with the pressure being uniform along the entire fiber at the given temperature. The refractive index can be approximated as $\sqrt{1+\chi}$, with~\cite{mccutcheon:2022aa}:
\begin{equation}\label{Eq:Refractive_index}
    \chi = N_{j} \frac{3\lambda_0^3 \gamma_0}{4\pi^2} \frac{i\gamma_0-2\delta}{\gamma_0^2 + 4\delta^2 +2|\Omega|^2},
\end{equation}
where  $\delta$ is the detuning and $N_j$ is the number of atoms in each cell of the fiber.  Since we are operating at low pressures, ranging from $10^{-8}$ to $10^{-7}$ bar (thus atomic density from $10^{11}$ to $10^{12}$), the refractive index of the gas is approximately 1, and the gas dispersion can be considered negligible. Additionally, the waveguide has anomalous dispersion, which can vary depending on the geometry of the hollow core fiber. However, given that the fiber length is relatively short at 50 mm, we assume the dispersion to be negligible for the presented data. Furthermore, in this regime, we neglect the effect of pressure broadening and incorporate its impact through the atom density. 

To accurately account for the nonlinearity in each segment of the fiber where the atoms are located, we divide the fiber into multiple segments. When the gas is at a low temperature, the atomic transitions can be well-approximated by a delta function due to negligible thermal broadening. However, at room temperature, thermal effects cause the atomic transitions to broaden, resulting in a Voigt profile. In this paper, we will discuss both scenarios.

For convenience, we transform Eqs.~(\ref{field_R_6}) into a retarded-time frame, $\tau =t-z/v_g$, where the reference frame propagates with the pulse center along the $z$-direction at velocity $v_g$~\cite{najafabadi:2024aa}. In this case, the pulse propagation follows:
\begin{align}
\left [ \frac{\partial}{\partial z}  + \left (\frac{1}{c}-\frac{1}{v_{g}} \right ) 
\frac{\partial}{\partial \tau} \right ] \Omega(\tau, z)  =  -\frac{1}{2} \kappa \Omega(\tau, z) \nonumber 
+ \sum_{\alpha=1} \sum_{\beta=1} \frac{G_{\alpha}}{c} \int\rho_{\alpha}(z,\omega) A_{\alpha} R^{-}_{\alpha \beta}(\tau, z,\omega)d\omega + F^{\Omega}(\tau, z).
\end{align}
and the atomic variables at each point along the fiber evolve in $\tau$.  We assume the atoms are initially distributed in the ground state, thus the polarization vectors $R^{\pm}_{\alpha \beta}$ and the population inversion $R^{3}_{\alpha \beta }$ vectors are $R^{+}_{\alpha \beta } = R^{-}_{\alpha \beta}=0$ and $R^{3}_{\alpha \beta }=-1/2$,  and the input coherent field, has the soliton shape  
\begin{equation}
    {\Omega}(0, t)  = 2 A \cosh^{-1} [A(\tau-\tau_{0})] \exp\left \{ i[ \delta \tau+\phi(0)] \right\} , 
\end{equation}
where $2A$ is the pulse amplitude, $\tau$ the pulse timing in the retarded time frame, at $z=0$, and $\delta$ is detuning of pulse. 

{
When the atomic distribution follows a Voigt profile\cite{garcia:2006aa}, $f_{\text{Voigt}}(\omega)$, the nonlinearity terms are calculated as:
\begin{equation}\label{eq:nonlinear}
\sum_{\alpha} \sum_{\beta} G_{\alpha}\int {R}_{\alpha \beta }(z,\Delta \omega_{\text{Voigt}}) \rho_{\text{$\alpha$}}(z) f_{\text{Voigt}}(\omega) d\omega ,
\end{equation}
where $\Delta \omega_{\text{Voigt}}$ is FWHM of the atomic Voigt profile and 
\begin{equation}
\Delta \omega_{\text{Voigt}} = \frac{\Delta \omega_{\text{L}}}{2} +\sqrt{{\Delta \omega_{\text{D}}}^2 + 0.2166~\Delta \omega_{\text{L}}^2},
\end{equation}
where $\Delta \omega_{\text{L}}$ is the Lorenzian FWHM and
\begin{equation}
    \Delta \omega_{\text{D}}^2 = \bigg ( \frac{4\pi}{\lambda_{0}}\bigg ) \sqrt{\frac{2\log(2)k_b T}{m_{i}}}
\end{equation}
is the Doppler FWHM at the temperature $T$. $\lambda_{0}$ is the wavelength of the selected transition, and $m_i$ is the isotope reduced atomic mass.}

The standard way of detecting squeezing is via homodyne detection with a local oscillator of amplitude  $f_{\mathrm{LO}}$ that we take as normalized $\int \left| {f}_{\mathrm{LO}} (t) \right|^2dt =1$.  {The local oscillator in our setup is assumed to have the same shape as the original pulse. This ensures optimal overlap for homodyne detection, as the local oscillator must match the temporal and spectral profile of the signal pulse to measure its quadratures accurately.} The normalized variances at the point $z$ along the fiber is.
\begin{align}
    \hat{M}(z)=\int_{-\infty}^{\infty} \left \{ {f}_{\mathrm{LO}}(t) \hat{\Omega}^{\dag}(t,z) e^{i\theta} + {f}^\dag_{\mathrm{LO}}(t) \hat{\Omega}(t,z) e^{-i\theta}\,\right \}  dt .
\end{align}
The corresponding squeezing ratio is
\begin{align}
    S(z)= \frac{\mathrm{Var} [\hat{M}(z)]}{\left.\mathrm{Var}  [\hat{M} ]\right|_{\rm coh}} \, ,
\end{align}
where $\mathrm{Var} [\hat{M}(z) ]= {\langle \hat{M}^2(z)\rangle} - {\langle \hat{M}(z) \rangle}^2$.

To calculate the variance with a $+{P}$ simulation, we need to express it in terms of normally ordered correlations:
{
\begin{align}
    \hat{M}^{2}\simeq \, :\hat{M}^{2}: + \frac{4\sum_{\alpha} G_{\alpha}}{v_g}  \, ,
\end{align}
 where we have used the approximate equal-space commutation relation $ [ \hat{\Omega}(t,z) , \hat{\Omega}^{\dag}(t^{'}, z)]\simeq 4\sum_{\alpha} G_{\alpha} \delta(t-t^{'})/v_g$. The squeezing ratio is then
\begin{align}
\label{Sq_Q}
    S = 1+ \frac{v_g\mathrm{Var}_{+P}
   [\hat{M}]}{4 \sum_{\alpha} G_{\alpha} }\, ,
\end{align}
where $\mathrm{Var}_{+P} \equiv {~\langle :\hat{M}^{2}: \rangle} - {\langle \hat{M} \rangle}^{2}$.}
%%%%%%%%%%%%%%%%%%%%%%%%%%%%%%%%%%%%%%%

\section{Results}
\label{sec:res}
\subsection{Low-temperature squeezing (273~K)}
 We calculate the quadrature squeezing from the variances of $\Omega$ using {Eq.~\eqref{Sq_Q}} after the optical pulse starts propagating in the PCF. By adjusting the phase of the local oscillator ${f}_\mathrm{LO}$, we can measure the squeezing at different phases at different propagation length.

The suitable transition for SIT is $\ce{^{3}_{}D_2}$ $\rightarrow$ $\ce{^{3}_{}P_2}$, with  $\lambda_0=365.5$~nm. To excite this transition, we consider a pulse with a duration of $\tau c/\lambda_0=0.44$, where $\tau=4$~fs, and can excite the $\ce{^{3}_{}S_1}$ $\rightarrow$ $\ce{^{3}_{}P_2}$ transition that is accounted in the simulation. { It is important to note that the pulse duration must be sufficiently short to minimize the effects of damping and dephasing. In this case, the linewidth of a femtosecond  pulse is much broader than the atomic linewidth, and significant implications for the dynamics can arise.

In the regime of SIT, the essential trick is to increase the peak intensity of the laser pulse to the point where one period of the Rabi frequency equals the pulse duration. In other words, by increasing the intensity, the atomic linewidth is power broadened~\cite{Allen:2012aa} sufficiently to match the wide spectrum of the laser pulse. In this regime, the atomic resonance no longer strictly corresponds to the narrow linewidth of the two-level atom model; instead, the atomic response becomes broader, and the detuning of the fs pulse relative to the atomic transition also increases: $\Delta \nu = \sqrt{\Delta \nu_L^2 +\Omega^2}$,  where $\Delta \nu_L$ specifies the natural linewidth of the atoms, and the Rabi frequency is proportional to the square root of the intensity.  
}

%%%%%%%%%%%%%%%%%%%%%%%%%%%%%%%% Compare the pulse duration on damping process 
\begin{figure}[b]
    \centering
    \includegraphics[width=0.8\columnwidth]{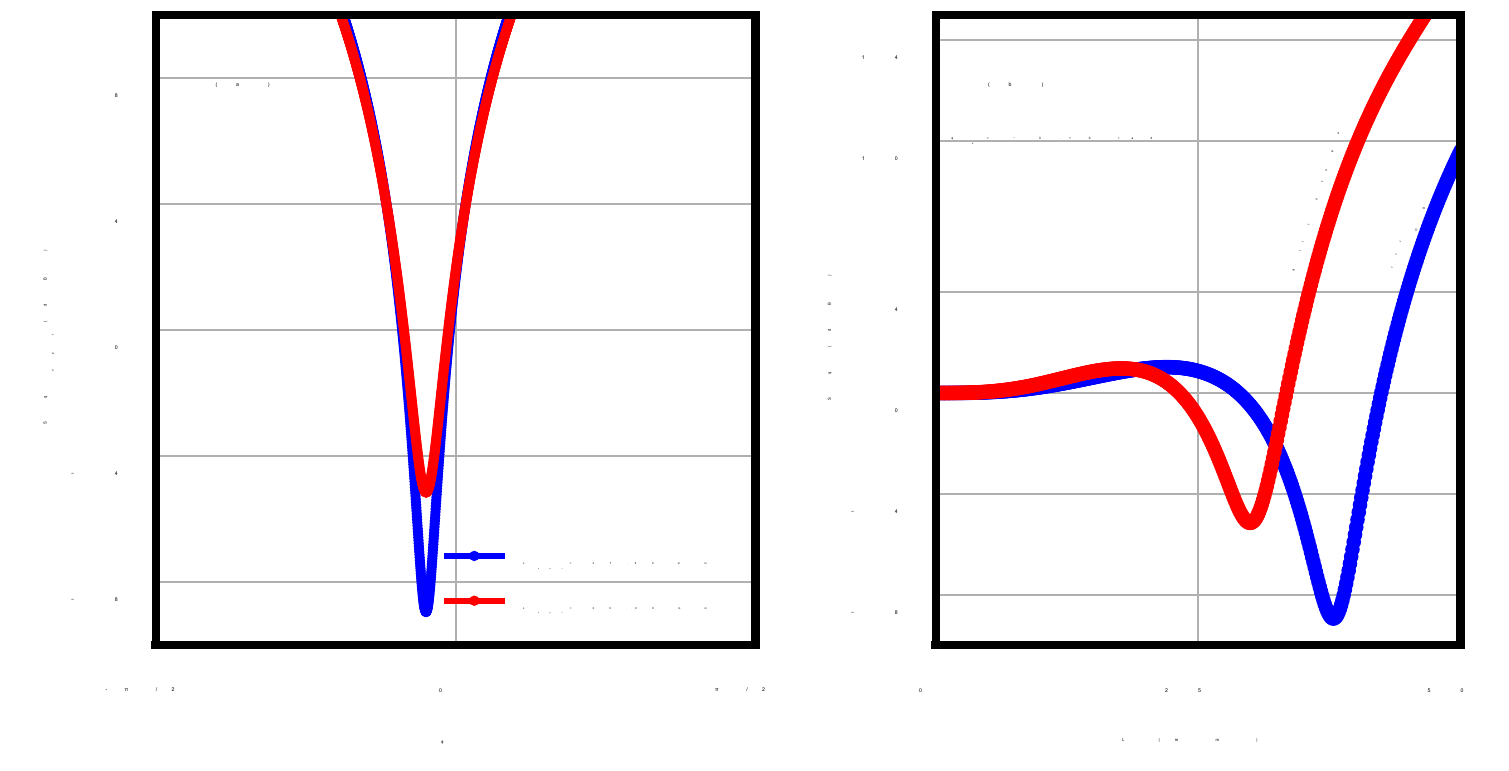}
    \caption{(a) The optimum squeezing as a function of the local oscillator phase and (b) the squeezing at the optimum angle as a function of fiber length for the isotope $\ce{^{202}_{}Hg}$  (blue curve) and for a gas including all isotopes (red curve).
    The pulse duration is considered to be 4 fs, and the atom pressure corresponds to the vapor pressure of mercury at T=273~K.}
    \label{fig:damping_effect_pulse_duration}
\end{figure}
%%%%%%%%%%%%%%%%%%%%%%%%%%

We consider two scenarios. First, we focus on the bosonic isotope $\ce{^{202}_{}Hg}$, as it has the highest natural abundance, and then include the fermionic isotopes with their hyperfine structures shown in Fig~\ref{fig:Energy_diagram}. In the second scenario, we utilize each isotope abundance to determine its corresponding atomic density within the whole sample.  Subsequently, we compute the polarization vectors $R^{\pm}$, $R^{3}$ for each given isotope separately. 

%%%%%%%%%%%%%%%%%%%%%%%%%
\begin{figure}[t]
    \centering
    \includegraphics[width=0.8\columnwidth]{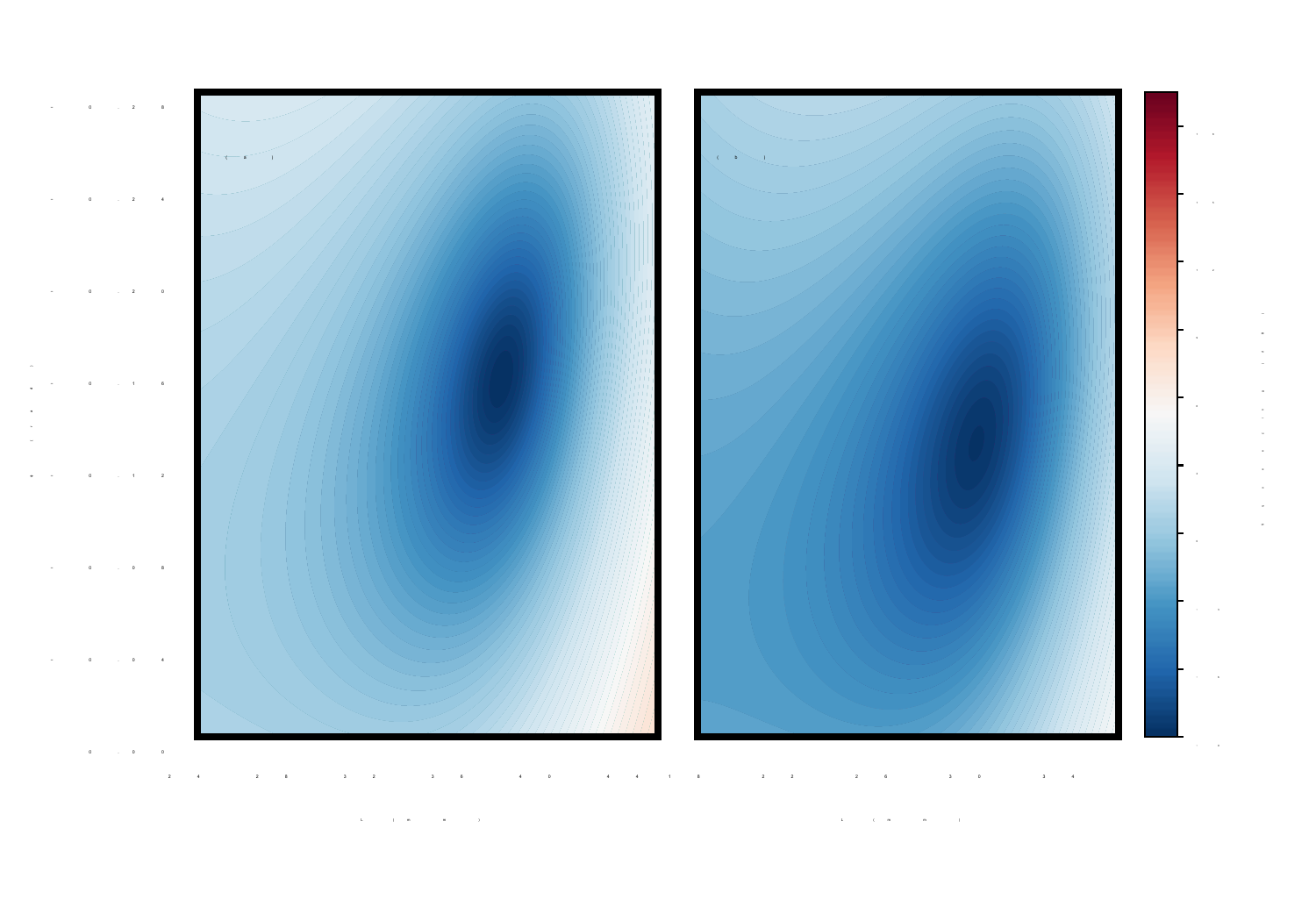}
    \caption{A two-dimensional color map showing the squeezing angle versus propagation length at T=273~K. (a) for isotope 202 and (b) for a gas including all the isotopes. }
    \label{fig:Heat_map_damping_effect_pulse_duration}
\end{figure}
%%%%%%%%%%%%%%%%%%%%%%%%%%%%%%%%%%%%%%%%%%%%%%%%%%%%%%%%%%%%%%%%%
%%%%%%%%%%%%%%%%%%%%%%%%%%%%%%%%%%%%%%%%%%%%%%%%%%%%%%%%%%%%% effect of detuning at T=273
\begin{figure}[t]
    \centering
    \includegraphics[width=0.8\columnwidth]{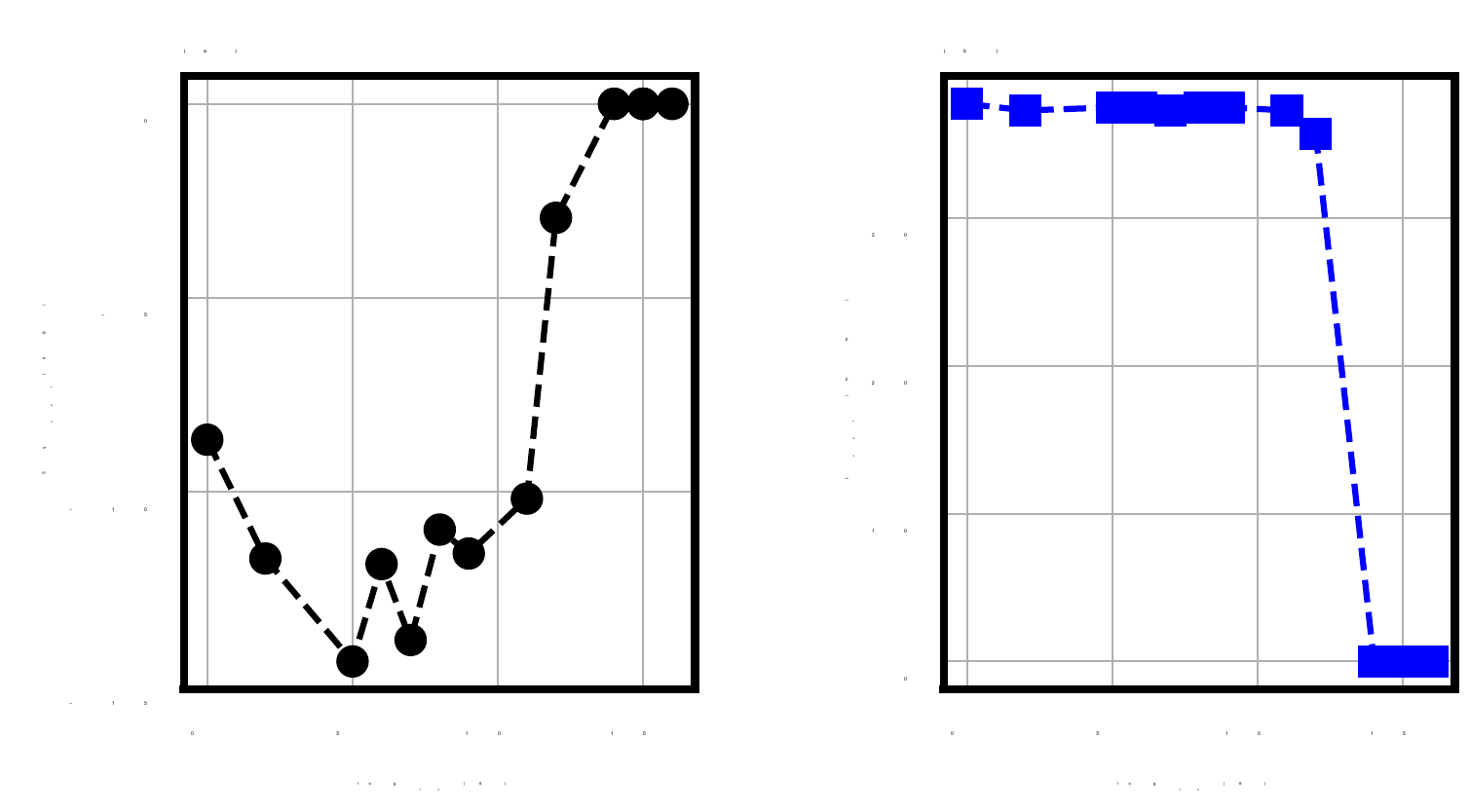}
    \caption{(a) Optimum squeezing as a function of the detuning, and (b) the detection length for the optimum squeezing as a function of detuning. The range of detuning changes from 0 to $16/\tau$. The pulse duration is $\tau=4$~fs and atomic vapour pressure is $\mathrm{P}=272 \times 10^{-8}$~bar at $T=273$–K.  A small offset of 1 is added to the detuning values before applying $\log_{10}$. }
    \label{fig:opt_Sq_L_dtuning}
\end{figure}
%%%%%%%%%%%%%%%%%%%%%%%%%%%%%%%%%%%%%%%%%%%%%%%%%%%%%%%%%%%%%%%%%%

{ For the bosonic isotopes ($I=0$), we compute the Bloch vectors only for the $6\ce{^{3}_{}D_{3}} \rightarrow 6\ce{^{3}_{}P_{2}}$ and $\ce{^{3}_{}S_1}$ $\rightarrow$ $\ce{^{3}_{} P_2}$ transitions. In contrast, for fermionic isotopes ($I\neq0$), the Bloch vector must be computed for each dipole-allowed transition in the hyperfine structure shown in Fig~\ref{fig:Energy_diagram}. These Bloch vectores are evaluated for each of these transitions separately and included in the nonlinear term. The nonlinear atom-field interaction term is then computed at a given position $z$ where the atoms are located .}

The atomic density is increased by raising the pressure through heating the gas. At $T = 273$~K, the gas pressure is taken to be 272 $\times 10^{-8}$ bar, which corresponds to an atomic number of approximately $14 \times 10^{11}$ in the entire fiber, and {atomic distribution forms a Voigt profile~\cite{garcia:2006aa}}.

Figure~\ref{fig:damping_effect_pulse_duration} (a) compares the optimum squeezing as a function of the detection phase of the local oscillator at 273 K. The maximum squeezing is achieved when a local oscillator phase is slightly away from zero ($\phi=-0.16$ rad). As the phase moves towards $\pi/2$ or $-\pi/2$, the optimal squeezing diminishes and eventually becomes zero. {The optimum squeezing is calculated as the maximum squeezing over a certain fiber length}.

Figure~\ref{fig:damping_effect_pulse_duration} (b) depicts the squeezing along the fiber at $\phi=-0.16$ rad. The fermionic isotopes, $\ce{^{199}_{}Hg}$ and $\ce{^{201}_{}Hg}$, encounter 30$\%$ of the mercury sample. The strength of the hyperfine transitions of these involved isotopes is of the order of GHz. As it is clear from the Fig.~\ref{fig:damping_effect_pulse_duration}, once these isotopes are considered in the simulation, the computed squeezing is slightly suppressed.  

{Figure~\ref{fig:Heat_map_damping_effect_pulse_duration} shows a two-dimensional color map of the squeezing angle versus propagation length. As observed, the squeezing angle changes with variations in the fiber length.}

\begin{figure}[t]
    \centering
    \includegraphics[width=0.8\columnwidth]{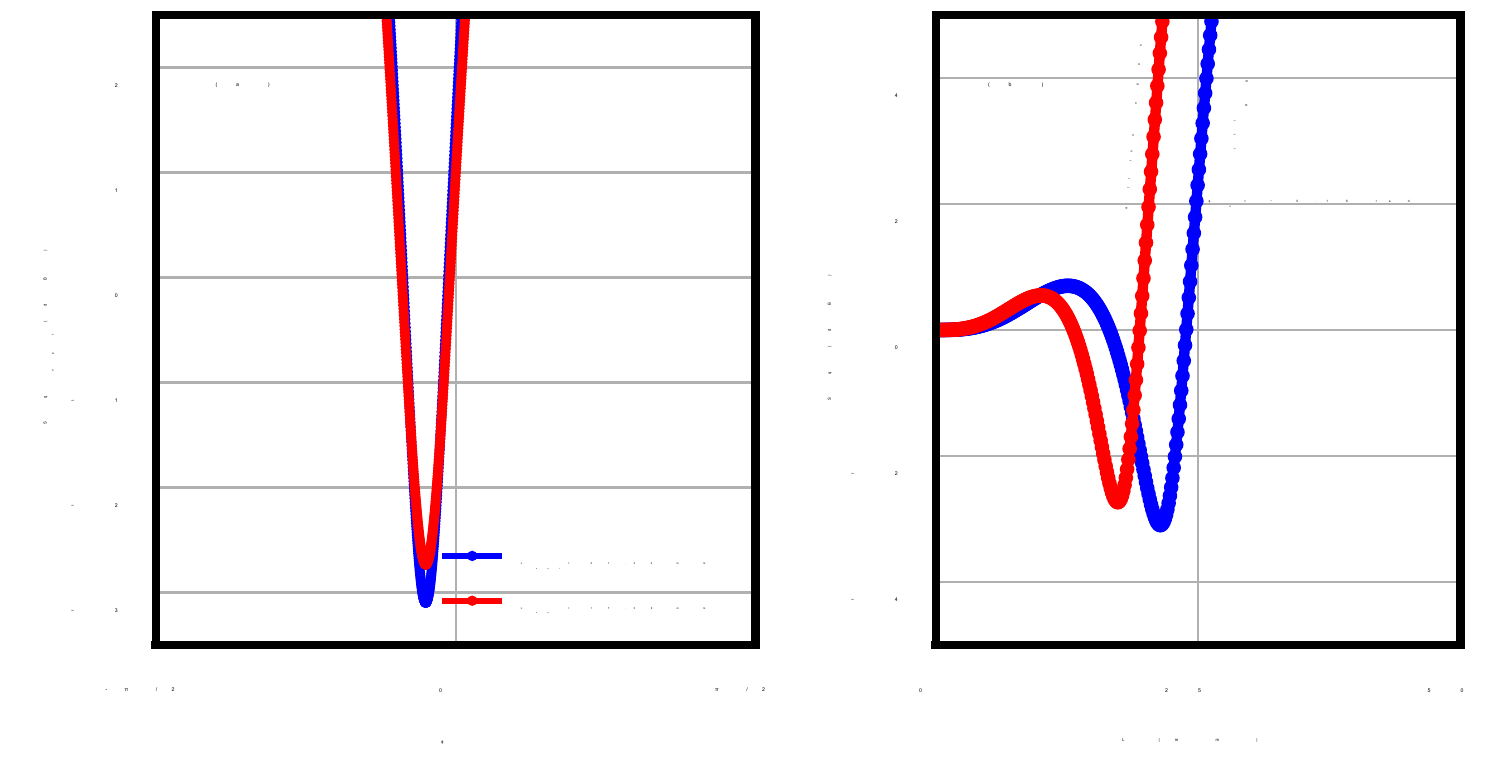}
    \caption{(a)Optimum squeezing as a function of the local oscillator phase and (b) squeezing at the optimum angle as a function of fiber length for the isotope $\ce{^{202}_{}Hg}$  (blue curve) and for a gas including all isotopes (red curve).
    The pulse duration is considered to be 4 fs, and the atom pressure corresponds to the vapor pressure of mercury at $T=293K$.
    }
    \label{fig:Sq_L_N_Room_Tem}
\end{figure}
%%%%%%%%%%%%%%%%%%%%%%%%%%%%%%%%%%% Heat map plot
\begin{figure}[t]
    \centering
    \includegraphics[width=0.8\columnwidth]{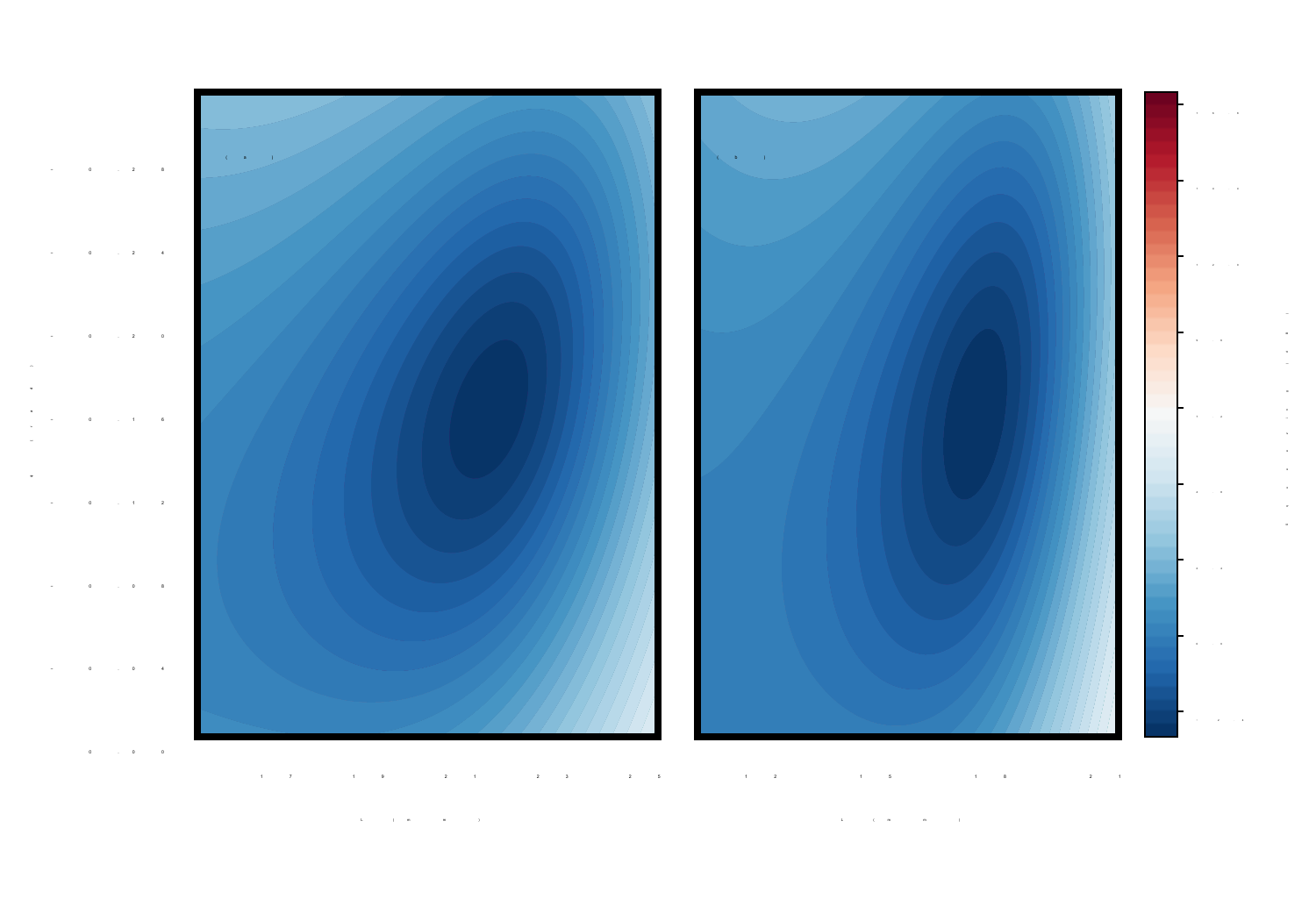}
       \caption{A two-dimensional color map  showing the squeezing angle versus propagation length at T=293 K. (a) for  isotope 202 and (b) for a gas including all isotopes}
    \label{fig:HeatMap_T_293}
\end{figure}
%%%%%%%%%%%%%%%%%%%%%%%%%%%%%%%%%%%%%%%%%%%%%%%%%

Figure~\ref{fig:opt_Sq_L_dtuning}(a) shows the optimum squeezing over the entire fiber length and the detection phases as a function of detuning (ranges from 0 to $16/\tau$). The black circle points show the optimum squeezing as a function of detuning.  Clearly, for a slightly detuned optical pulse ($\delta < 10^{-3}/\tau$ ), the squeezing exceeds its value compared to a pulse at resonance. Beyond this point, as the detuning increases, the squeezing gradually reduces to zero. This emphasizes that SIT squeezing at 273 K is achievable when the pulse and atoms are either at resonance or even slightly detuned. { 

The apparent irregularities in the data arise from random fluctuations, reflecting sampling error. This error becomes more significant at extreme squeezing values. Increasing the sample size  could improve the smoothness of the data. In this work, we used approximately 10,000 to 12,000 samples for each detuning value. }

Figure~\ref{fig:opt_Sq_L_dtuning}(b) illustrates the detection length $L_{\mathrm{opt}}$ corresponding to the optimal squeezing. With increasing detuning, the detection length remains unchanged for pulses on resonance, except when the squeezing reduces to  zero.

\subsection{Squeezing at room temperature}

%%%%%%%%%%%%%%%%%%%%%%%%%%%%%%%%%%%%%%%%%%%%%%%%%%%%%%%%%%%%%%%% effect of dtuning at T = 293 %%%%%%%%%%%%%%%%%%%%%%%%%%%%%%%%%%
\begin{figure}[t]
    \centering
    \includegraphics[width=0.9\columnwidth]{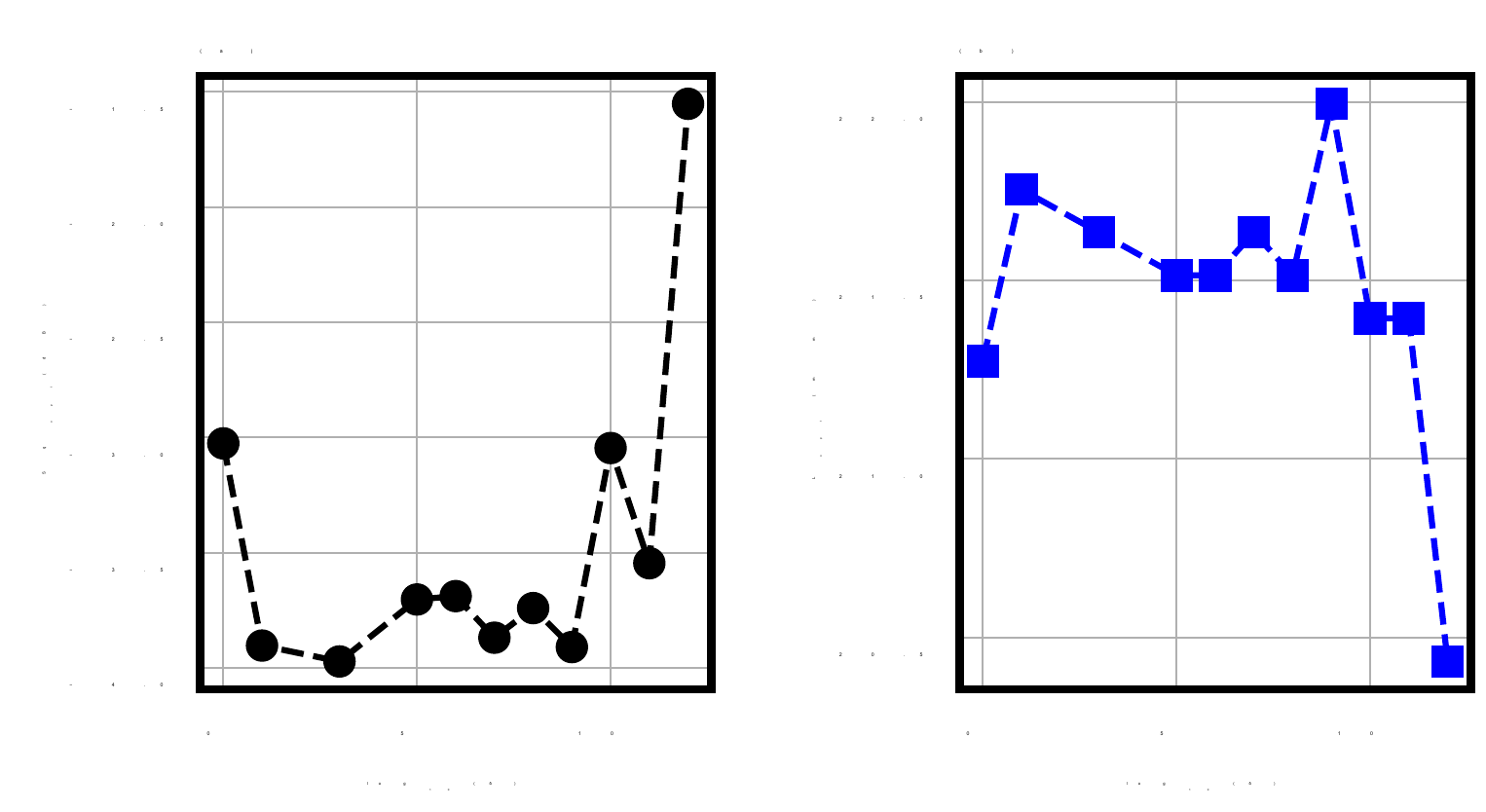}
       \caption{(a) Optimum squeezing as a function of detuning, and (b) the detection length  of the optimum squeezing. The pulse duration is $\tau=4 \mathrm{fs}$ and atomic vapour pressure is $\mathrm{P}=8.89 \times 10^{-6}$ bar at $T=293$ K.  A small offset of 1 is added to the detuning values before applying $\log_{10}$.}
    \label{fig:opt_Sq_L_dtuning_T_293}
\end{figure}
%%%%%%%%%%%%%%%%%%%%%%%%%%%%%%%%%%%%%%%%%%%%%%%%%%%%%%%%%%%%%%%%%%
%%%%%%%%%%%%%%%%%%%%%%%%%%%%%%%%%%%%%%%%%%%%%%%%%%%%%%%%%%%%%%%%% show the optimum squeezing for two pulse duration with two atomic density 
\begin{figure}[t]
    \centering
    \includegraphics[width=0.9\columnwidth]{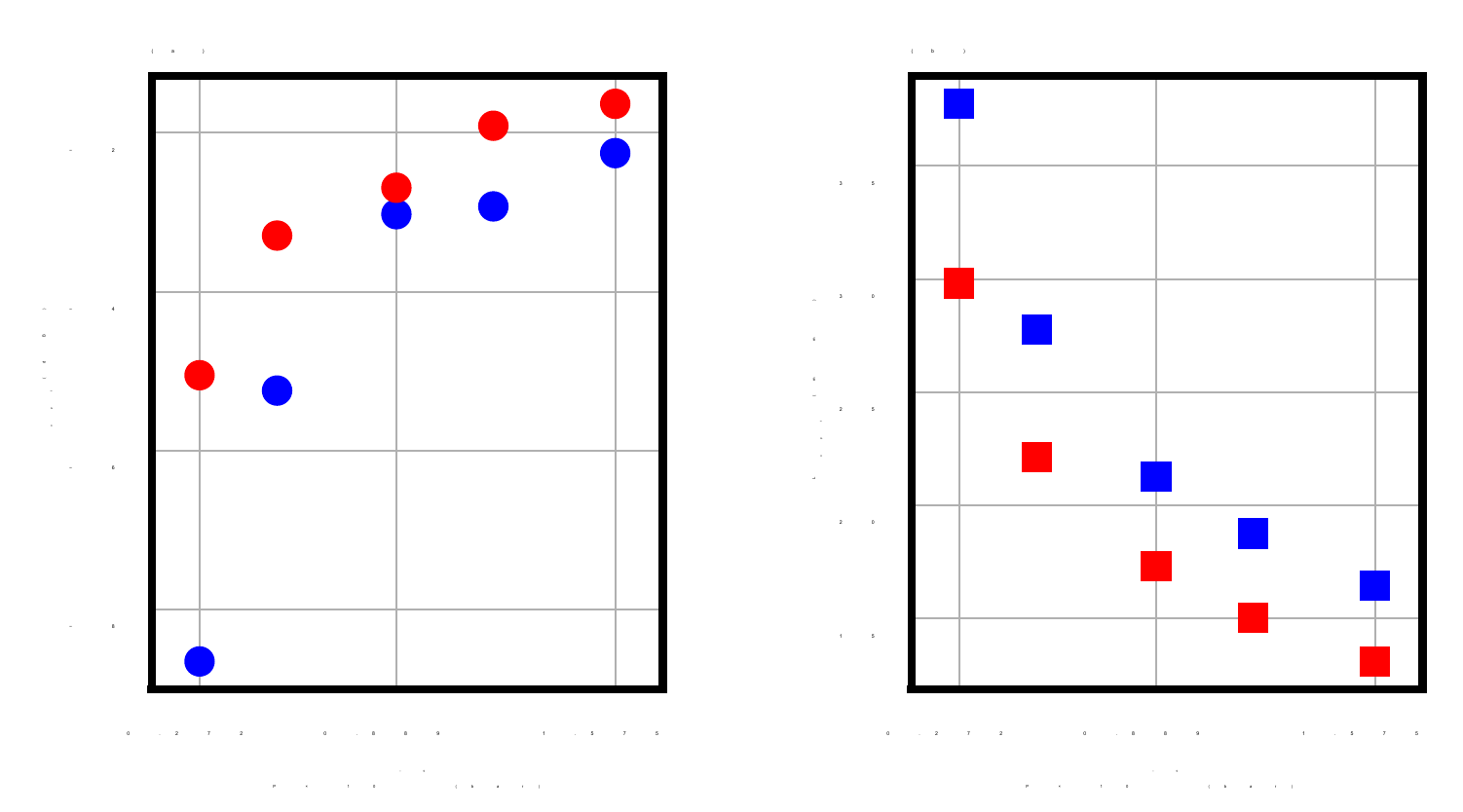}
    \caption{
    (a) Optimum squeezing as a function of pressure, blue circle circles (only $\ce{^{202}_{}Hg} $), the red circles, includes all isotopes. (b) the detection length for optimum squeezing as a function of pressure. } 
    \label{fig:optmimum_sq_fiber_Tem}
\end{figure}
%%%%%%%%%%%%%%%%%%%%%%%%%%%%%%%%%%%%%%%%%%%%%%%%%%Conclusion%%%%%%%%%%%%%%%%%%%%%%%%%%%%%%%%%%%%%%%%%%%%%%%%%%%
{In this section, we focus on room temperature to explore the extent of squeezing achievable under these conditions and to determine the optimal squeezing angle.
}
%%%%%%%%%%%%%%%%%%%%%%%%%%%%%%%%%%%%Results %%%%%%%%%%%%%%%%%%%%%%%%%%%%%%%%%%%%%%%%%%%%%%%%%%%%%%%%%%%%%%

Figure~\ref{fig:Sq_L_N_Room_Tem} compares the squeezing at { a specific fiber length} at 293K using a coherent pulse on resonance with a duration $\tau=4$fs.{ Additionally, Fig.~\ref{fig:HeatMap_T_293} shows the squeezing angle as a function of propagation length, which varies with changes in the fiber length.}

For sufficiently short pulse duration, it is possible to achieve squeezing at room temperature, although the effect is not significant. Higher temperatures enhance damping and dephasing, leading to a reduction in squeezing. 

%%%%%%%%%%%%%%%%%%%%%%%%%%%%%%%%%%%%%%%%%%%%%%%%%%%%%%%%%%%%%%%%%%%%%%%% detuning

As it is depicted in Fig.~\ref{fig:opt_Sq_L_dtuning_T_293}, the detuning leads to only 1dB  squeezing when the atomic absorption line shape follows a Voigt profile at a temperature of 293 K.

%%%%%%%%%%%%%%%%%%%%%%%%% opt squeezing as a function of pressure %%%%%%%%%%%%%%%%%
Figure~\ref{fig:optmimum_sq_fiber_Tem} (a) compares the optimum squeezing as a function of pressure for isotope $\ce{^{202}_{}Hg}$ (blue points) and a sample including all isotopes (red points). Figure~\ref{fig:optmimum_sq_fiber_Tem}(b) shows the corresponding detection length as a function of vapor pressure. As the pressure increases, the squeezing diminishes, and the detection length shortens. Notably, the inclusion of all isotopes in the sample has little effect on the squeezing compared to the case where only the isotope $\ce{^{202}{}Hg}$ is considered.

\section{Concluding remarks}
\label{sec:Conc}
{We have studied the feasibility of quantum squeezing using mercury vapor gas in a hollow-core PCF. { In our model, we assume that the local oscillator pulse shape matches the original pulse to ensure optimal overlap for homodyne detection, considering only atoms on-resonance. However, under specific conditions, such as interactions with off-resonant atoms or the presence of Kerr  interactions, the squeezing structure of the pulse can develop multimode characteristics. This phenomenon has been investigated in previous studies~\cite{hosaka:2016aa,ng:2023aa,zhang:2016aa}, which explore multimode temporal and spatial structures arising from nonlinear propagation and atomic interactions.}

Our findings show that the best squeezing is achievable at 273 K since damping suppresses squeezing.  Mercury proves to be a promising candidate for SIT squeezing; however, careful consideration must be given to the selection of the pulse duration and wavelength. To fully excite this transition, we consider a pulse with a duration of $\tau = 4$ fs {  representing the shortest pulse duration for a given bandwidth at the carrier frequency $\omega_0$. While longer pulses, such as those with a duration of approximately 80~fs, can also support squeezing in our model, the squeezing decreases with longer pulse durations due to damping and dephasing effects. However, employing laser cooling techniques to reduce the longitudinal motion of the atoms can effectively suppress these effects. Under such conditions, the primary sources of noise are significantly minimized, allowing for higher levels of squeezing to be achieved.

The selected transition occurs at a wavelength of $\lambda = 365.5$ nm. To achieve this wavelength, it would typically require frequency conversion, such as third harmonic generation, from an Optical Parametric Oscillator (OPO) output.} Furthermore, proper tuning of the pulse and optimization of other parameters, such as gas pressure and fiber length, are crucial to maximize the squeezing effect.

\begin{backmatter}
\bmsection{Funding}
Agencia Estatal de Investigaci\'on (PID2021-127781NB-100).

\bmsection{Acknowledgments}
We acknowledge discussions with U. Vogl. , T. Dirmeier and C. Genes. 

\bmsection{Disclosures}
The authors declare no conflicts of interest.

\bmsection{Data availability} 
Data underlying the results presented in this paper are not publicly available at this time but may be obtained from the authors upon reasonable request.

\end{backmatter}

%\bibliography{SIT}

\end{document}